\documentclass[reprint,amsmath,amssymb,aps,prl
,superscriptaddress]{revtex4-2}

\usepackage{graphicx}
\usepackage{physics}
\usepackage{hyperref}
\usepackage{caption}
\usepackage{xcolor}

\newcommand{\mytexttilde}{\raisebox{0.5ex}{\texttildelow}}

\makeatletter
\renewcommand{\section}[2][]{}
\renewcommand{\subsection}[2][]{}
\renewcommand{\subsubsection}[2][]{}
\makeatother

\newcommand{\SIQSE}{\affiliation{1}{Shenzhen Institute for Quantum Science and Engineering, Southern University of Science and Technology, Shenzhen, Guangdong, China}}
\newcommand{\DPHY}{\affiliation{2}{Department of Physics, Southern University of Science and Technology, Shenzhen, Guangdong, China}}
\newcommand{\IQA}{\affiliation{3}{International Quantum Academy, Shenzhen, Guangdong, China}}
\newcommand{\GDKL}{\affiliation{4}{Guangdong Provincial Key Laboratory of Quantum Science and Engineering, Southern University of Science and Technology, Shenzhen, Guangdong, China}}
\newcommand{\HFNL}{\affiliation{5}{Shenzhen Branch, Hefei National Laboratory, Shenzhen 518048, China}}

\begin{document}

\title{Quantum dynamics of microwave photons in synthetic frequency dimension}

\author{Zheshu Xie}
%\email{xiezs2022@mail.sustech.edu.cn}
%\email{12232912@mail.sustech.edu.cn}
\thanks{These authors contributed equally.}
\affiliation{\SIQSE}\affiliation{\IQA}\affiliation{\GDKL}

\author{Luojia Wang}
%\email{ljwang@sjtu.edu.cn}
\thanks{These authors contributed equally.}
\affiliation{State Key Laboratory of Photonics and Communications, School of Physics and Astronomy, Shanghai Jiao Tong University, Shanghai 200240, China}

\author{Jiawei Qiu}
\thanks{These authors contributed equally.}
\affiliation{\SIQSE}\affiliation{\IQA}\affiliation{\GDKL}

\author{Libo Zhang}
\affiliation{\SIQSE}\affiliation{\IQA}\affiliation{\GDKL}

\author{Yuxuan Zhou}
\affiliation{\IQA}

\author{Ziyu Tao}
% \email{taoziyu@iqasz.cn}
\affiliation{\IQA}\affiliation{\GDKL}

\author{Wenhui Huang}
% \email{huangwh2022@mail.sustech.edu.cn}
\affiliation{\SIQSE}\affiliation{\IQA}\affiliation{\GDKL}

\author{Yongqi Liang}
% \email{liangyq2023@mail.sustech.edu.cn}
\affiliation{\SIQSE}\affiliation{\IQA}\affiliation{\GDKL}

\author{Jiajian Zhang}
% \email{zhangjj2020@mail.sustech.edu.cn}
\affiliation{\SIQSE}\affiliation{\IQA}\affiliation{\GDKL}

\author{Yuanzhen Chen}
\affiliation{\DPHY}\affiliation{\SIQSE}\affiliation{\GDKL}

\author{Song Liu}
\affiliation{\SIQSE}\affiliation{\IQA}\affiliation{\GDKL}\affiliation{\HFNL}

\author{Jingjing Niu}
\affiliation{\IQA}\affiliation{\HFNL}

\author{Yang Liu}
\email{liuyang\_jlu2007@126.com}
\affiliation{\IQA}

\author{Youpeng Zhong}
\email{zhongyp@sustech.edu.cn}
\affiliation{\SIQSE}\affiliation{\IQA}\affiliation{\GDKL}\affiliation{\HFNL}

\author{Luqi Yuan}
\email{yuanluqi@sjtu.edu.cn}
\affiliation{State Key Laboratory of Photonics and Communications, School of Physics and Astronomy, Shanghai Jiao Tong University, Shanghai 200240, China}

\author{Dapeng Yu}
%\email{yudapeng@iqasz.cn}
\affiliation{\SIQSE}\affiliation{\IQA}\affiliation{\GDKL}\affiliation{\HFNL}
\date{\today}

\begin{abstract}
Synthetic frequency dimension offers a powerful approach to simulate lattice models and control photon dynamics. However, extending this concept into the quantum regime, particularly at the single-photon level, has remained challenging in photonic platforms. 
% due to limited photon control efficiency and weak interaction strength. 
Here, we demonstrate quantum-state initialization and detection of single-photon evolutions within a synthetic frequency lattice by integrating a superconducting qubit with a 16-meter aluminum coaxial cable. A tunable superconducting quantum interference device (SQUID)-based modulator is employed to synthesize lattice couplings and artificial gauge fields. We observe single-photon quantum random walks and Bloch oscillations, as well as nonadiabatic, unidirectional frequency conversion under rapid temporal modulation of the lattice Hamiltonian, together with band-structure measurements. The lattice connectivity can be readily reconfigured to construct higher-dimensional lattices using multiple drive tones. Our results establish superconducting quantum circuits as a versatile platform for programmable Hamiltonians and extensible synthetic lattices with flexible single-photon control.
\end{abstract}

\maketitle

\section{Introduction.}
\vspace{1.5em} 
{\noindent\large\bfseries Introduction} 
\vspace{0.8em}\\
\begin{figure*}[t!]
    \centering
    \vspace{-10pt}
    \includegraphics[width=0.7\textwidth]{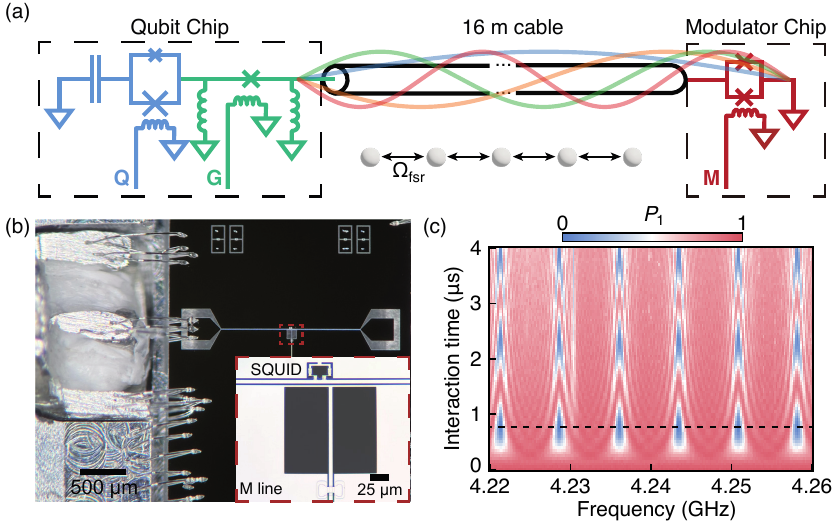}
    \caption{
    \justifying
    Experimental setup.
    (a) Schematic of the setup. A flux-tunable qubit (blue) is connected to a 16-meter low-loss superconducting cable via a gmon tunable coupler (green). The opposite end of the cable is connected to a modulator (red) before terminated to ground. 
    (b) Optical micrograph of the modulator. The cable extends to the chip and is connected via direct wirebonds. Zoom in reveals the detailed structure the modulator SQUID. The open-end port on the right is short compared to the mode wavelengths and can be neglected.
    (c) Exciteted state population of qubit in the single-photon vacuum-Rabi oscillations with the cable modes. The JC coupling $\kappa/2\pi=0.36$~MHz is kept weak for individually addressing the modes. The black dashed line indicates the full swap time.
    }
    \label{fig1}
\end{figure*}
Physical properties are profoundly shaped by the system's dimensionality, influencing phenomena such as quantum Hall effects~\cite{zhang2001four,tang2019three,zhang2019quantum,42zilberberg2018photonic}, topological phases~\cite{slobozhanyuk2017three,jia2019observation,cooper2019topological,maczewsky2020synthesizing}, localization~\cite{schwartz2007transport,conti2008dynamic,yamilov2023anderson} and quantum entanglement in high dimensions~\cite{14erhard2020advances,hu2020efficient}.
Many experimental simulations have been realized in spatial dimensions of different quantum systems like Rydberg atoms~\cite{bernien2017probing,de2019observation,semeghini2021probing,scholl2021quantum}, microwave cavities~\cite{fitzpatrick2017observation,kollar2019hyperbolic} and superconducting qubits~\cite{roushan2017spectroscopic,ma2019dissipatively}.
However, the quest to simulate higher-dimensional quantum phenomena has been constrained by the spatial limitations of existing experimental platforms. 
This challenge has driven a surge of interest in synthetic dimensions~\cite{1yuan2018synthetic,30ozawa2019topological,lustig2021topological,ehrhardt2023perspective,hazzard2023synthetic,arguello2024synthetic}, 
which not only provides additional dimensionality to Hamiltonian but also offers a cost-effective and flexible simulation means through new methods of control and measurement.

Photonic systems have been at the forefront of synthetic dimension explorations due to their simple design, flexible tunability, and potential applications~\cite{1yuan2018synthetic}. 
For example, the uniformly spaced modes of a resonant ring form discrete lattice sites that can be connected by external dynamic modulation, providing a foundation for compact construction of synthetic frequency dimensions~\cite{3dutt2020single,heckelmann2023quantum,cheng2025non}. 
However, most of the works on synthetic dimensions in photonics are inclined to classical systems, where realizing single-photon quantum dynamics are hindered by large losses, limited efficiency of photon production and detection~\cite{wang2019towards,tomm2021bright,charaev2023single}, and weak photon-photon interactions~\cite{couteau2023applications}. 
Recent studies using superconducting waveguides have established key techniques for controllable inter-mode coupling and semi-classical simulations of synthetic lattices~\cite{2lee2020propagation,hung2021quantum,busnaina2024quantum,ahrens2025synthetic}. These monolithic platforms reduce losses and enable experiments with fewer photons but still face challenges for single-photon dynamics.
Driven by the quest for large-scale quantum computers, low-loss superconducting cables suitable for interfacing with superconducting qubit chips have been developed recently~\cite{63kimble2008quantum, Kurpiers2018, Axline2018, Niu2023, Storz2023, 26qiu2023deterministic, chang2020remote, malekakhlagh2024enhanced}, opening new opportunities for exploring quantum dynamics in synthetic frequency dimensions.

In this work, we realize quantum simulations in a synthetic frequency lattice within a superconducting-circuit platform operating at the single-photon quantum regime~\cite{clerk2020hybrid}. By integrating a superconducting qubit with a 16-meter aluminum cable and employing a tunable SQUID-based modulator, we achieve flexible control over the system Hamiltonian and directly observe the quantum evolution of individual photons.
In this platform, single-photon states in a single site or as a wave packet can be prepared, and the Hamiltonian can be abruptly changed in time. These capabilities extend the accessible regime of synthetic frequency dimensions beyond prior photonic implementations~\cite{englebert2023bloch,javid2023chip,dikopoltsev2025collective}. 
Moreover, by introducing long-range inter-mode couplings, effective gauge fluxes can be engineered to fold the frequency dimension, enabling synthetic constructions of higher-dimensional models~\cite{senanian2023programmable,wang2024chip} in the single-photon regime.

\section{Experimental setup.}
\vspace{1.5em} 
{\noindent\large\bfseries Results} 
\vspace{0.8em}\\
The experimental setup is illustrated in Fig.~\ref{fig1}(a). 
A superconducting qubit chip is connected to one end of a 16-meter aluminum superconducting coaxial cable, with the other end linked to a SQUID-based modulator, both with low-loss direct wirebonds~\cite{29zhong2021deterministic} providing galvanic connections. The gap between the cable and chips is minimized, and multiple short wirebonds are used in parallel to minimize impedance mismatch.
The cable supports a series of evenly spaced standing-wave modes whose angular frequencies are given by:
\begin{equation}
\omega_m=\omega_0+m \Omega_{\mathrm{fsr}},
\end{equation}
where $m$ denotes the mode index, $\omega_0$ is the base frequency, and $\Omega_{\mathrm{fsr}}/2\pi=7.33$~MHz is the free spectral range (FSR). The small FSR results from the long cable, which can be readily extended without the yield limitations that typically restrict the length of on-chip coplanar waveguides.
Such a small FSR allows access to more than 30 adjacent modes within a few hundred megahertz, providing sufficient lattice sites in frequency lattices for simulating complex or higher-dimensional models within the experimentally accessible operational window. 
Moreover, the cable exhibits relatively high coherence ($T_1\sim29.1~\mathrm{\mu s}$, $T_2\sim57.9~\mathrm{\mu s}$), making it well-suited for studying quantum dynamics in the resulting synthetic frequency lattices. 

A transmon qubit~\cite{16koch2007charge} is used as a single-photon source and as a detector of the mode states.
We use the set of standing-wave modes near the qubit frequency of 4.32~GHz, corresponding to mode numbers $m\in[576,608]$. 
For simplicity, the mode closest to the qubit frequency is relabeled as $m = 0$, and all other modes are indexed relative to this reference.
Coupling between the qubit and the cable is mediated by a gmon tunable coupler~\cite{chen2014qubit}. 
The Jaynes-Cummings (JC) coupling strength $\kappa_m$ between the qubit and the $m$-th mode scales as $\sqrt{\omega_m}$~\cite{27zhong2019violating} and gives rise to vacuum-Rabi oscillations at weak coupling.
Across the \mytexttilde230~MHz operational bandwidth of our experiment, the coupling strengths remain nearly constant and can be simply denoted as $|\kappa|/2\pi$, with a tuning range from $0$ to $7.4$~MHz afforded by the gmon tunable coupler.

Inter-mode coupling is introduced via the SQUID-based modulator (Fig.~\ref{fig1}(b))~\cite{zakka2011quantum,sandbo2018generating}. 
The SQUID functions as a flux-controlled inductance that can be tuned from a finite value to effectively infinity, sweeping the electrical boundary of the cable modes from near short to near open, thereby shifting the frequencies of all modes.
A microwave drive on the SQUID bias line periodically modulates the mode frequencies and enables photon hopping between the modes. The Hamiltonian is:

\begin{align}
    H=\,&\omega_q\sigma^{\dagger}\sigma + \sum_m \omega_m a_m^{\dagger} a_m + \sum_m \kappa_m (\sigma^{\dagger} a_{m}+a_{m}^{\dagger} \sigma) \nonumber
    \\&+\sum_{m,n,l} (-1)^{m+n}2 g_l \cos (\Omega_l t+\phi_l)\left(a_m^{\dagger} a_{n}+a_{n}^{\dagger} a_m\right),  \label{Hamiltonian}
\end{align}
where we set $\hbar = 1$ for convenience. $\omega_q$ is the qubit angular frequency, $a_m$ and $\sigma$ are the annihilation operators for the $m$-th resonant mode and the qubit respectively. 
% When focusing solely on the photon dynamics in the cable, the qubit terms can be neglected by tuning $\kappa_m$ to zero.
When $\kappa_m$ is tuned to zero, the qubit decouples and its terms can be omitted.
The last term introduces a time-dependent coupling between different modes.
Each drive tone $l$ is characterized by frequency $\Omega_l$, phase $\phi_l$, and associated hopping strength $g_l$. %The alternating sign reflects the standing-wave parity and guarantees the correct gauge structure after rotating-wave approximation.
Moving to the rotating frame $\widetilde{c}_m = e^{i(\omega_m+\!m\Delta)t}\,a_m$ and performing the rotating-wave approximation, we obtain the effective Hamiltonian (ignoring the qubit part):
\begin{align}
\widetilde{H}_{\mathrm{RWA}}&= -\sum_m m\Delta\widetilde{c}_m^\dagger\widetilde{c}_m\nonumber\\ 
&+\sum_{l,m}(-1)^{l}g_l(e^{i\phi_l}\widetilde{c}_m^\dagger\widetilde{c}_{m+l}+e^{-i\phi_l}\widetilde{c}_{m+l}\widetilde{c}_m^\dagger).
\label{RWA2}
\end{align}
Usually we tune the driving frequencies to $\Omega_l = l(\Omega_{\mathrm{fsr}}+\Delta)$ to selectively couple modes spaced by $l\,\Omega_{\mathrm{fsr}}$. A small detuning $\Delta\ll \Omega_{\mathrm{fsr}}$ imposes a constant effective force in the synthetic frequency lattice, leading to Bloch oscillations and related dynamics~\cite{longhi2005dynamic,7yuan2016bloch,6hu2020realization}.

\section{Dynamics under single-site excitation.}
\begin{figure}[ht]
    \centering    
    \includegraphics[width=0.45\textwidth]{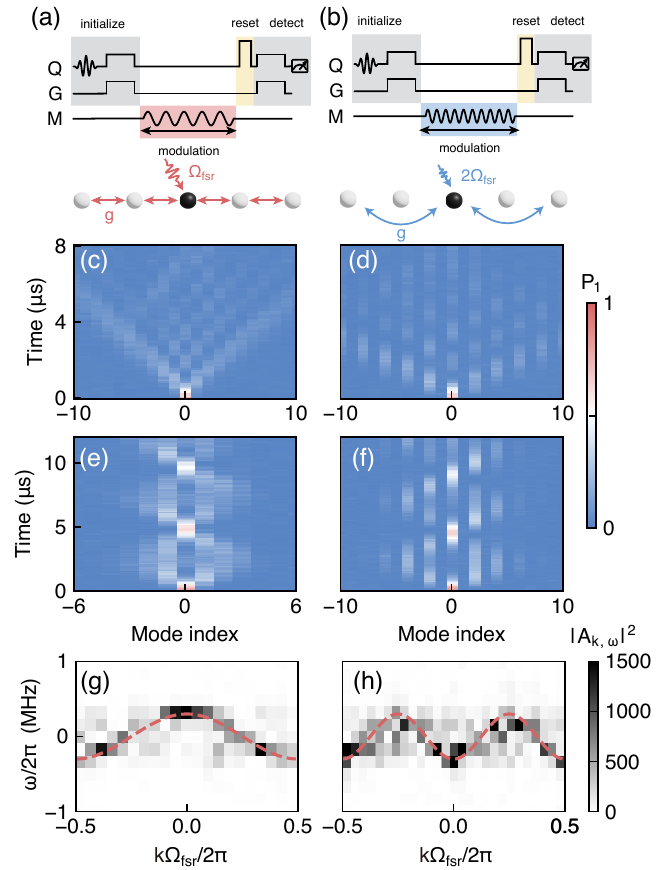}
    \caption{
    \justifying
    Dynamics of single-site excitations in the frequency dimension.
    (a),(b) Schematic of the pulse sequence, showing the steps for initial state preparation, modulation application, residual photon resetting, and final state measurement. 
    (c),(d) Time evolution for the on-resonance modulation case with drive frequencies of $\Omega/2\pi = 7.33$~MHz and $14.66$~MHz, respectively.
    (e),(f) Time evolution for the off-resonance modulation case with drive frequencies of $\Omega/2\pi = 7.13$~MHz and $14.46$~MHz, respectively. 
    (g),(h) Energy band structures for a 1D tight-binding model with nearest- and next-nearest-neighbor coupling at $\phi=\pi$.
    }
    \label{fig2}
\end{figure}

We first observe the evolution of a single-site initial state, which, after applying a Fourier transform, becomes broadly distributed across the entire $k$-space. This wide distribution samples the global band properties. 
We excite the qubit and weakly couple it to the cable at $\kappa/2\pi = 0.36$~MHz $\ll \Omega_{\mathrm{fsr}}$, so individual modes are addressed independently, as shown in Fig.~\ref{fig1}(c). 
The interaction time is set to the first Rabi minimum to swap the excitation completely into the selected cable mode, after which the qubit-cable coupling is turned off.
The single photon then evolves within the cable under the applied modulations. After a given delay, we read out a chosen mode by reversing the swap and measuring the qubit excited-state population $P_1$. Note that any residual qubit excitation is rapidly dissipated through the lossy readout line before the mode detection.
Repeating this procedure over all modes reconstructs the single-photon population dynamics under the target Hamiltonian. The pulse sequence is shown in Fig.~\ref{fig2}(a) and (b)  for two drive frequencies.
With $\Omega=\Omega_{\mathrm{fsr}}$, nearest-neighbor hopping dominates, while with $\Omega=2\Omega_{\mathrm{fsr}}$, next-nearest-neighbor hopping dominates. In both cases we observe the corresponding quantum random walks~\cite{aharonov1993quantum} in frequency space (Fig.~\ref{fig2}(c),(d)).
Adding a detuning $\Delta=-0.2$~MHz yields Bloch oscillations with period 4.9~$\mathrm{\mu s}$ (Fig.~\ref{fig2}(e),(f)), consistent with $T_B=2\pi/|\Delta|=5~\mathrm{\mu s}$ in theory.

% To directly resolve the band structure of the synthetic lattice, 
Observing the single-photon dynamics requires long coherence time of the modes, which precludes the conventional spectral method for detecting the energy bands. 
Thus we exploit the qubit and adapt a protocol from superconducting qubit arrays~\cite{roushan2017spectroscopic,55morvan2022formation,56neill2021accurately,62xiang2023simulating}. 
The single-photon wavefunction in the synthetic frequency lattice is
\begin{equation}
|\psi\rangle = C_{\mathrm{vac}} |0\rangle + \sum_{m} C_m |1_m\rangle,\label{wavepacket_state}
\end{equation} 
where $|0\rangle$ is vacuum and $|1_m\rangle$ denotes one photon in mode $m$. We obtain the mode-resolved quadratures by swapping each mode back to the qubit and measuring the expectation values of $\hat{X}_m$ and $\hat{Y}_m$. A two-dimensional Fourier transform of $\psi(m,t)$ yields the dispersion relation $\omega(k)$, thereby mapping the band structure as shown in Fig.~2(g), (h), in good agreement with theory. More details can be found in the Supplementary Material.
% ~\cite{Xie2025_SM}.

% that usually has been observed in photonic counterparts~\cite{morandotti1999experimental,englebert2023bloch,javid2023chip}.

\begin{figure}[ht]
    \centering
    \includegraphics[width=0.45\textwidth]{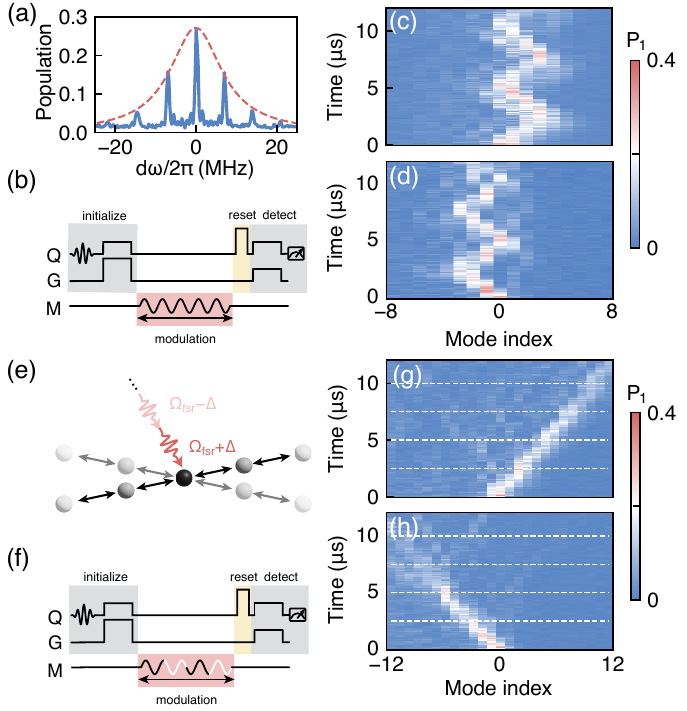}
    \caption{
    \justifying
     Wave-packet excitation propagation.
    (a) Initial distribution of the photon wave packet, achieved by tuning the gmon coupler to a strong-coupling regime. The red dotted line represents the fitted Lorentzian envelope.
    (b) Schematic of the pulse sequence.
    (c),(d) Evolution of the wave packet under single-tone driving with $\Omega/2\pi = 7.13$~MHz and $7.53$~MHz, corresponding to detunings $\Delta/2\pi=-0.2$~MHz and $\Delta/2\pi=0.2$~MHz.
    (e),(f) Schematic of modulation with a periodically changing frequency and the corresponding pulse sequence. 
    (g),(h) Evolution of the wave packet under periodic detuning reversal, starting at $\Omega/2\pi = 7.13$~MHz and $7.53$~MHz, where the sign of $\Delta$ is reversed every half-cycle (2.5~$\mathrm{\mu s}$).
    }
    \label{fig3}
\end{figure}

\section{Dynamics under wave-packet excitation.}
A single-site excitation is spatially localized and hence occupies the full Brillouin zone, producing bidirectional spreading. To probe local band properties, we prepare a spectrally narrow wave packet. We tune the gmon coupler to a stronger coupling, $\kappa/2\pi = 4$~MHz, comparable to $\Omega_{\mathrm{fsr}}$, so the qubit couples coherently to multiple standing-wave modes, realizing a multimode JC interaction. Starting from the excited qubit, the emitted single photon populates a superposition across discrete modes, yielding a state of the form in Eq.~\ref{wavepacket_state} and a population distribution shown in Fig.~\ref{fig3}(a). Fitting the distribution with a Lorentzian (see Supplementary Material) gives a full width at half maximum of $31$~MHz and a peak value of $0.27$ at zero detuning.
% The coupling strength between the qubit and cable modes follows a Lorentzian distribution as a function of the frequency detuning $d\omega=\omega_m - \omega_q$:
% \begin{equation}
% P(\dd\omega) = P_0 \, 
% \frac{(\Gamma/2)^2}{
% (\dd\omega)^2 + (\Gamma/2)^2},
% \label{eq:Lorentzian}
% \end{equation}
% where $\Gamma$ is the spectral linewidth and \red{$P_0$} is a normalization constant. 
% The experimentally measured mode population $|\beta_m|^2$ is plotted in Fig.~3(a).
% The best fit yields $\Gamma/2\pi = 31.3~\mathrm{MHz}$ and $P_0 = 0.272$.
% Compared with a coherent-state pulse, this single-photon wave packet carries a single quantum of excitation with a controllable spectral--temporal shape, enabling direct exploration of non-classical photon emission and reabsorption dynamics in the non-Markovian multimode regime.
% The numerical simulation of wave-packet distribution in real space and $k$-space is shown in the Supplementary Information~\cite{Xie2025_SM}.
Following a similar pulse sequence as in the single-site experiment (Fig.~\ref{fig3}(b)), the packet evolution exhibits asymmetric Bloch oscillations, as shown in Fig.~\ref{fig3}(c). This asymmetry arises because the initial wave-packet excitation coherently populates a finite region of the energy band around $k=0$, unlike the single-site state.
Although the group velocity at the packet center is zero initially (see Fig.~\ref{fig2}(g)), the effective force immediately shifts the packet into the region with positive group velocity $v_g=\dd\omega/\dd k$ ($k\Omega_{\mathrm{fsr}} \in (-\pi, 0)$) producing rightward motion near $t=0~\mathrm{\mu s}$ in Fig.~\ref{fig3}(c). 
At half a period ($t=2.5~\mathrm{\mu s}$), the packet center approaches $k\Omega_{\mathrm{fsr}}=-\pi$, and continued forcing reverses the direction ($k\Omega_{\mathrm{fsr}} \in (0, \pi)$). 
The oscillation direction flips if we reverse the detuning, e.g. to $\Delta/2\pi = 0.2$~MHz, as presented in Fig.~\ref{fig3}(d). 

We further demonstrate non-adiabatic control by periodically reversing the synthetic force (detuning $\Delta$, Fig.~\ref{fig3}(g)) every 2.5~$\mathrm{\mu s}$, exactly half a Bloch oscillation period.
By doing so, we can induce unidirectional transport of the photon in frequency space, as shown in Figs.~\ref{fig3}(g) and (h).
Chiral transfer in real-space waveguides was realized recently~\cite{almanakly2025deterministic}. In frequency space, unidirectional motion was theoretically predicted~\cite{7yuan2016bloch} but had not been observed due to the challenge of rapid temporal modulation. These temporal operations may also connect to emerging time-varying photonics~\cite{galiffi2023broadband,jaffray2025spatio}.

\begin{figure}[t!]
    \centering
    \includegraphics[width=0.45\textwidth]{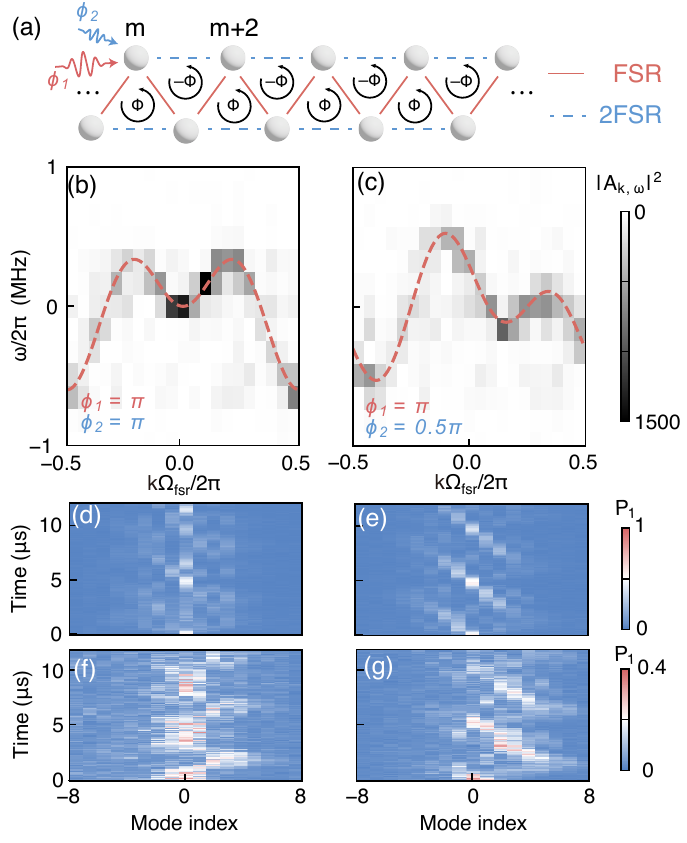}
    \caption{
    \justifying
    Single-photon dynamics with a synthetic gauge field.
    (a) The equivalent lattice is a triangular chain, with each plaquette characterized by a phase $\Phi$.
    (b),(c) Energy bands for double driving with phases $\phi_1 =  \pi$, $\phi_2 = \pi$ and $0.5\pi$.
    (d),(e) Time evolution of a single-site excitation under simultaneous driving at $7.13$~MHz and $14.26$~MHz, with phases $\phi_1 =  \pi$, $\phi_2 = \pi$ and $0.5\pi$. 
    (f),(g) Time evolution of a wave-packet state under the same driving conditions as (d) and (e).
    \label{fig4}
    }
\end{figure}

\section{Effective magnetic flux.}
Thus far, our demonstration is constrained to a one-dimensional synthetic lattice. It is also possible to fold the lattice and simulate higher-dimensional models in the quantum limit. As a proof of concept, we implement double driving that simultaneously couples nearest-neighbor and next-nearest-neighbor modes. The phases on the two driving signals create a photonic gauge potential~\cite{vepsalainen2020simulating,tusnin2020nonlinear,hung2021quantum} with flux $\Phi = 2\phi_1 - \phi_2$ per triangular plaquette (Fig.~\ref{fig4}(a)). 
Lattices with more complex connectivities and dimensionalities can also be engineered by adding more drive tones.
This synthetic gauge field gives rise to a more complex band structure, as verified in Fig.~\ref{fig4}(b),(c) for $\Phi = \pi, 1.5\pi$, where the latter shows band asymmetry. 
We again study Bloch oscillations with driving frequencies set as $\Omega_1 = \Omega_{\mathrm{fsr}}+\Delta$ and $\Omega_2=2\Omega_1$. 
The band asymmetry directly manifests in the dynamics: Fig.~\ref{fig4}(e) shows asymmetric evolution compared with the otherwise symmetric case in Fig.~\ref{fig4}(d) for the single-site excitation. For wave packets, both cases (see Fig.~\ref{fig4}(f, g)) are asymmetric, with stronger distortion in Fig.~\ref{fig4}(g). 
% The present single-cable geometry introduces non-zero gauge flux in the single-photon limit, thereby breaking time-reversal symmetry and induces asymmetric band structure in Fig.~\ref{fig4}(c)~\cite{34dutt2019experimental}. However, due to the single-band nature, a single cable here does not simulate a topologically nontrivial model. Nevertheless, the precise phase control in Fig.~\ref{fig4} demonstrates the essential prerequisite for realizing topological phases and the long wavelength of microwave by constructing complex synthetic lattices with multiple cables or via device engineering in future. Incorporating longer-range couplings would enable effective dimensional folding and twisted boundary conditions~\cite{yuan2018syntheticPRB,21schwartz2013laser,senanian2023programmable}.
While the present single-cable system allows for nonzero gauge flux and hence breaks time-reversal symmetry and produces nonreciprocal bands~\cite{34dutt2019experimental} such as Fig.~\ref{fig4}(c), its effectively single-band nature prevents the realization of a topological phase. The flexible phase control demonstrated here nevertheless establishes the essential ingredients for future topological implementations. In particular, the long microwave wavelength and the design flexibility of superconducting circuits make it practical to combine multiple cables or engineer multi-band structures, enabling uniform synthetic flux, longer-range couplings, and twisted boundary conditions. These advances would open the route to synthetic dimensional folding and topological models~\cite{yuan2018syntheticPRB,21schwartz2013laser,senanian2023programmable}.

\section{Conclusion.}
\vspace{1.5em}
{\noindent\large\bfseries Discussion and conclusion} 
\vspace{0.8em}\\
We have realized single-photon quantum dynamics in a synthetic frequency lattice using a superconducting qubit, a 16~m low-loss superconducting cable, and a SQUID modulator.
Unlike previous studies on synthetic frequency dimensions in fiber-based~\cite{3dutt2020single,englebert2023bloch,senanian2023programmable,cheng2025non} or monolithic platforms~\cite{heckelmann2023quantum,6hu2020realization,wang2024chip,dikopoltsev2025collective} that remain primarily in classical or semi-classical regimes, our work operates fully at the quantum level, %targeting single-photon processes. We realize synthetic lattices in the frequency space of an low-loss superconducting cable with modulated inter-site couplings,
enabling quantum coherent control and detection of individual photons.
Notably, our qubit-assisted state readout enables single-shot measurement of quantum states of individual photons, in contrast to ensemble-averaged field detection methods, such as steady-state measurement~\cite{ahrens2025synthetic} and site-resolved transient measurement~\cite{2lee2020propagation}.
This platform further provides direct access to programmable Hamiltonians that incorporate various effective gauge fields and fluxes.  These capabilities mark a key step toward quantum simulations in synthetic dimensions. Looking ahead, precise control over the quantum state of each mode would allow for the preparation of more complex non-gaussian states and enable the study of multi-photon interactions. Furthermore, by coupling multiple cables with independently tunable interactions, the platform can be extended to implement synthetic gauge fields with nontrivial topological properties.

\section{Acknowledgments.}
\vspace{1.5em}
{\noindent\large\bfseries Acknowledgments} 
\vspace{0.8em}\\
This work was supported by the Science, Technology and Innovation Commission of Shenzhen Municipality (KQTD20210811090049034), the National Natural Science Foundation of China (12174178, 12204228, 12374474, 12122407, 12204304, 12504582), National Key Research and Development Program of China (No. 2023YFA1407200), the Innovation Program for Quantum Science and Technology (2021ZD0301703).

\vspace{1.5em} 
{\noindent\large\bfseries Competing interests} 
\vspace{0.8em}\\
The authors declare that they have no conflict of interest.

\bibliographystyle{apsrev4-2}
\bibliography{ref}

\end{document}